\documentclass[12pt]{article}
\usepackage{graphicx}
\begin{document}

%%-move to normal A4-%%
\hoffset = -0.7truecm
\voffset = -1.9truecm

\title{\bf
Comments on the Monopole-Antimonopole Pair Solutions\footnote{Paper to be submitted for publication}}

\author{
\textit{\textbf {Rosy Teh and K.M. Wong}}\\
\textit{{\normalsize School of Physics}}\\
\textit{{\normalsize Universiti Sains Malaysia}}\\
\textit{{\normalsize 11800 USM Penang}}}

\date{January 2010}
\maketitle

\begin{abstract}
Recently, the monopole-antimonopole pair and monopole-antimonopole chain solutions are solved with internal space coordinate system of $\theta$-winding number $m$ greater than one. 
However, we notice that it is also possible to solve these solutions numerically in terms of $\theta$-winding number $m=1$ instead. When $m=1$, the exact asymptotic solutions at small and large distances are  parameterized by a single integer parameter $s$. Here we once again study the monopole-antimonopole pair solution of the SU(2) Yang-Mills-Higgs theory which belongs to the topological trivial sector numerically in its new form. This solution with $\theta$-winding and $\phi$-winding number one is parameterized by $s=0$ at small $r$ and $s=1$ at large $r$.
\end{abstract}

%%-main body of paper-%%
%%-self numbering sections-%%

\section{Introduction}
The SU(2) Yang-Mills-Higgs (YMH) field theory in $3+1$ dimensions, with the Higgs field in the adjoint representation are known to possess a large varieties of magnetic monopole configurations. 
The first finite energy monopole solution is the 't Hooft-Polyakov monopole solution \cite{kn:1} which is invariant under a U(1) subgroup of the local SU(2) gauge group.
It has non zero Higgs mass and self-interaction and is a numerical solution. Later Prasad and Sommerfield \cite{kn:2} gave the closed form version of the 't Hooft-Polyakov monopole in the BPS limit.
The YMH field theory with a unit magnetic charge and finite energy is spherically symmetric \cite{kn:1}-\cite{kn:2}. However multimonopole configurations with magnetic charges greater than unity cannot possess spherical symmetry \cite{kn:3} but at most axial symmetry \cite{kn:4}.  

So far exact monopole and multimonopoles solutions  \cite{kn:2}, \cite{kn:4} existed only in the Bogomol'nyi-Prasad-Sommerfield (BPS) limit. Outside this limit, when the Higgs field potential is non-vanishing, only numerical solutions are known. 
We have also shown that the ansatz of Ref.\cite{kn:5} possesses more exact multimonopole-antimonopole configurations in the BPS limit. 

The axially symmetric monopole-antimonopole pair (MAP) of Kleihaus and Kunz \cite{kn:6}, and the monopole-antimonopole chain (MAC) of Kleihaus et al. \cite{kn:7}, \cite{kn:8} possess exact asymptotic solutions at both small and large distances. The connecting solutions at immediate distances are both numerical and non-BPS as they do not satisfy the Bogomol'nyi equation and are only solutions to the second order differential equations of motion.
Their solutions possess only axial symmetry. These MAP and MAC solutions are parameterized by  $\theta$-winding number $m ( > 1)$ and  $\phi$-winding number $n = 1$.

Here we notice that there are several different discrete sectors of the YMH vacuum as well as the Wu-Yang monopole system that can be represented by exact solutions when the $\theta$-winding number $m=1$ and parameterized by a single integer parameter $s$ \cite{kn:9}. The exact BPS one monopole solution of Ref. \cite{kn:2} connects the trivial YMH vacuum at small distances with the Wu-Yang monopole solution at large distances and it corresponds to the case when the parameter $s=0$ at both small and large distances.

We also note that at small $r$ both the MAP and MAC solutions correspond to the trivial vacuum with $m=1$ and $s=0$. However at large $r$, the MAP solutions tend to a different sector of the vacuum with the parameter $s=1, 2, 3,...$, $m=1$. Hence the MAP solutions correspond to a one MAP when at large $r$, $s=1$, a two MAP when at large $r$, $s=2$, and so on with zero net topological magnetic charge. On the contrary, the MAC solutions tend to a different sector of the Wu-Yang monopole system at large $r$ when $s=1, 2, 3,…$; $m=1$ and hence possess topological magnetic charge of one when $n=1$. In this case at large $r$, $s=1$ will give the M-A-M (monopole-antimonopole-monopole) solution, $s=2$ will give the M-A-M-A-M solution and so on \cite{kn:9}. 

We briefly review the SU(2) Yang-Mills-Higgs field theory in the next section. In section 3, we show that the  $\theta$-winding number of the MAP and MAC solutions can be reduced to one and a single integer parameter $s$ for both exact asymptotic solutions at large and small $r$. In other words, there exist an equivalent form of the solutions with normal $\theta$-winding number $m = 1$ and integer parameter $s$ for all the solutions of $\theta$-winding number $m$ \cite{kn:9}. In section 4, we present the BPS one monopole solution with $\theta$-winding number $m$ higher than one and we give the equivalent solutions when the $\theta$-winding $m$ is reduced to one. In section 5, we compute and present the numerical MAP solutions with $(n,m,s)=(1,1,1)$ and $(n,m,s)=(1,2,0)$ at large $r$ of different accuracies. Both the asymptotic conditions at large $r$ correspond to a pure gauge at spatial infinity and hence the system possess zero net magnetic charge. We connect the asymptotic solutions with the trivial vacuum at the origin to the vacuum solutions with $(n,m,s)=(1,1,1)$ and $(n,m,s)=(1,2,0)$ at large $r$ numerically by using mathematical softwares Maple and Matlab \cite{kn:10}. We end with some comments in section 6.

%%%%%%%%%%%%%%%%%%%%%%%%%%%%%%%The SU(2) YMH Theory%%%%%%%%%%%%%%%%%%%%%%%%%%%%%%%%%%%%%%%%%%%%%%%%%%%%%%%

\section{The SU(2) YMH Theory}
The SU(2) YMH Lagrangian in 3+1 dimensions with non vanishing Higgs potential is given by
\begin{equation}
{\cal L} = -\frac{1}{4}F^a_{\mu\nu} F^{a\mu\nu} - \frac{1}{2}D^\mu \Phi^a D_\mu \Phi^a - \frac{1}{4}\lambda(\Phi^a\Phi^a - \frac{\mu^2}{\lambda})^2. 
\label{eq.1}
\end{equation}

\noindent Here the Higgs field mass is $\mu$ and the strength of the Higgs potential is $\lambda$ which are constants. The vacuum expectation value of the Higgs field is $\xi=\mu/\sqrt{\lambda}$. The Lagrangian (\ref{eq.1}) is gauge invariant under the set of independent local SU(2) transformations at each space-time point.
The covariant derivative of the Higgs field and the gauge field strength tensor are given respectively by 
\begin{eqnarray}
D_{\mu}\Phi^{a} &=& \partial_{\mu} \Phi^{a} + g\epsilon^{abc} A^{b}_{\mu}\Phi^{c},\nonumber\\
F^a_{\mu\nu} &=& \partial_{\mu}A^a_\nu - \partial_{\nu}A^a_\mu + g\epsilon^{abc}A^b_{\mu}A^c_\nu.
\label{eq.2}
\end{eqnarray}
Since the gauge field coupling constant $g$ can be scaled away, we can set $g$ to one without any loss of generality. The metric used is $g_{\mu\nu} = (- + + +)$. The SU(2) internal group indices $a, b, c$ run from 1 to 3 and the spatial indices are $\mu, \nu, \alpha = 0, 1, 2$, and $3$ in Minkowski space.

The equations of motion that follow from the Lagrangian (\ref{eq.1}) are
\begin{eqnarray}
D^{\mu}F^a_{\mu\nu} &=& \partial^{\mu}F^a_{\mu\nu} + \epsilon^{abc}A^{b\mu}F^c_{\mu\nu} = \epsilon^{abc}\Phi^{b}D_{\nu}\Phi^c,\nonumber\\
D^{\mu}D_{\mu}\Phi^a &=& \lambda\Phi^a(\Phi^{b}\Phi^{b} - \frac{\mu^2}{\lambda}),
\label{eq.3}
\end{eqnarray}
and the Bogomol'nyi equation,
\begin{equation}
B^a_i \pm D_i \Phi^a = 0,
\label{eq.4}
\end{equation}
holds in the limit of vanishing $\mu$ and $\lambda$.
  
The tensor identified with the electromagnetic field, as proposed by 't Hooft \cite{kn:1} is
\begin{eqnarray}
F_{\mu\nu} &=& \hat{\Phi}^a F^a_{\mu\nu} - \epsilon^{abc}\hat{\Phi}^{a}D_{\mu}\hat{\Phi}^{b}D_{\nu}\hat{\Phi}^c,\nonumber\\
	&=& \partial_{\mu}A_\nu - \partial_{\nu}A_\mu - \epsilon^{abc}\hat{\Phi}^{a}\partial_{\mu}\hat{\Phi}^{b}\partial_{\nu}\hat{\Phi}^c,
\label{eq.5}
\end{eqnarray}

\noindent where $A_\mu = \hat{\Phi}^{a}A^a_\mu$, the Higgs unit vector, $\hat{\Phi}^a = \Phi^a/|\Phi|$, and the Higgs field magnitude $|\Phi| = \sqrt{\Phi^{a}\Phi^{a}}$. 
The Abelian electric field is $E_i = F_{0i}$, and the Abelian magnetic field is $B_i = -\frac{1}{2}\epsilon_{ijk}F_{jk}$. 
We would also like to write the Abelian 't Hooft magnetic field as
\begin{equation}
B_i = B_i^G + B_i^H,
\label{eq.6}
\end{equation} 
\begin{eqnarray}
\mbox{where} ~~~B_i^G = -\epsilon_{ijk}(\partial_{\mu}A_\nu - \partial_{\nu}A_\mu), ~~~
B_i^H = \epsilon_{ijk}\epsilon^{abc}\hat{\Phi}^{a}\partial_{\mu}\hat{\Phi}^{b}\partial_{\nu}\hat{\Phi}^c,
\label{eq.7}
\end{eqnarray}

\noindent which we refer to as the gauge part and the Higgs part of the 't Hooft magnetic field respectively. 

The topological magnetic current \cite{kn:11} is defined to be 
\begin{equation}
k_\mu = \frac{1}{8\pi}~\epsilon_{\mu\nu\rho\sigma}~\epsilon_{abc}~\partial^{\nu}\hat{\Phi}^{a}~\partial^{\rho}\hat{\Phi}^{b}~\partial^{\sigma}\hat{\Phi}^c,
\label{eq.8}
\end{equation}

\noindent which is also the topological current density of the system and the corresponding conserved topological magnetic charge is
\begin{eqnarray}
M & = & \int d^{3}x~k_0 = \frac{1}{8\pi}\int \epsilon_{ijk}\epsilon^{abc}\partial_{i}\left(\hat{\Phi}^{a}\partial_{j}\hat{\Phi}^{b}\partial_{k}\hat{\Phi}^{c}\right)d^{3}x \nonumber\\
& = & \frac{1}{8\pi}\oint d^{2}\sigma_{i}\left(\epsilon_{ijk}\epsilon^{abc}\hat{\Phi}^{a}\partial_{j}\hat{\Phi}^{b}\partial_{k}\hat{\Phi}^{c}\right)\nonumber\\
& = & \frac{1}{4\pi} \oint d^{2}\sigma_{i}~B_i^H. 
\label{eq.9}
\end{eqnarray}

\noindent As mentioned by Arafune et al. \cite{kn:12}, the magnetic charge $M$ is the total magnetic charge of the system if and only if the gauge field is non singular. If the gauge field is singular and carries Dirac string monopoles, then the total magnetic charge of the system is the total sum of the Dirac string monopoles and the monopoles carry by the Higgs field which is $M$.

Our work is restricted to the static case where $A^a_0=0$. Hence the conserved energy of the system for the static case reduces to

\[E=\int d^3 x \left\{\frac{1}{2}B^a_i B^a_i + \frac{1}{2}D^a\Phi_i D^a\Phi_i + \frac{1}{4}\lambda(\Phi^a\Phi^a - \frac{\mu^2}{\lambda})\right\}.\]                                               

Here $i$, $j$, $k$ which are the three space indices run from 1, 2, and 3. This energy vanishes when the gauge potential $A^a_\mu$ is zero or when it is a pure gauge, and when  $\Phi^a\Phi^a=\xi^2$ and $D_\mu\Phi^a=0$. In this paper, we consider the case with vanishing Higgs potential, $\lambda=0$. 

%%%%%%%%%%%%%%%%%%%%%%%%%%%% The Exact Asymptotic Solutions %%%%%%%%%%%%%%%%%%%%%%%%%%%%%%%
\section{The Exact Asymptotic Solutions}
The magnetic ansatz of Ref.\cite{kn:5}-\cite{kn:6} can be generalized to include unit vectors of the internal space with $\theta$-winding number $m>1$ and $\phi$-winding number $n>1$,
\begin{eqnarray}
A_i^a &=&  - \frac{1}{r}\psi_1(r, \theta) \hat{u}^{a}_\phi\hat{\theta}_i + \frac{1}{r}\psi_2(r, \theta)\hat{u}^{a}_\theta\hat{\phi}_i
+ \frac{1}{r}R_1(r, \theta)\hat{u}^{a}_\phi\hat{r}_i - \frac{1}{r}R_2(r, \theta)\hat{u}^{a}_r\hat{\phi}_i, \nonumber\\
\Phi^a &=& \frac{1}{r}\chi_1(r, \theta)~\hat{u}^a_r + \frac{1}{r}\chi_2(r, \theta)\hat{u}^a_\theta.
\label{eq.10}
\end{eqnarray}

\noindent The spatial spherical coordinate orthonormal unit vectors, $\hat{r}_i, \hat{\theta}_i$, and $\hat{\phi}_i$ are defined by 
\begin{eqnarray}
\hat{r}_i &=& \sin\theta ~\cos \phi ~\delta_{i1} + \sin\theta ~\sin \phi ~\delta_{i2} + \cos\theta~\delta_{i3},\nonumber\\
\hat{\theta}_i &=& \cos\theta ~\cos \phi ~\delta_{i1} + \cos\theta ~\sin \phi ~\delta_{i2} - \sin\theta ~\delta_{i3},\nonumber\\
\hat{\phi}_i &=& -\sin \phi ~\delta_{i1} + \cos \phi ~\delta_{i2}.
\label{eq.11}
\end{eqnarray}
They are the normal spherical coordinate orthonormal unit vectors with winding numbers $m=1$ and $n=1$. The isospin coordinate orthonormal unit vectors, $\hat{u}_r^a, \hat{u}_\theta^a$, and $\hat{u}_\phi^a$ are with integer winding numbers $m\geq 1$ and $n\geq 1$,
\begin{eqnarray}
\hat{u}_r^a &=& \sin m\theta ~\cos n\phi ~\delta_{1}^a + \sin m\theta ~\sin n\phi ~\delta_{2}^a + \cos m\theta~\delta_{3}^a,\nonumber\\
\hat{u}_\theta^a &=& \cos m\theta ~\cos n\phi ~\delta_{1}^a + \cos m\theta ~\sin n\phi ~\delta_{2}^a - \sin m\theta ~\delta_{3}^a,\nonumber\\
\hat{u}_\phi^a &=& -\sin n\phi ~\delta_{1}^a + \cos n\phi ~\delta_{2}^a.
\label{eq.12}
\end{eqnarray}

\noindent In this paper, MAP solutions refer to M-A-M-A-.....-M-A (MAPs) chain lying on the $z$-axis and MAC solutions refer to M-A-M-A-.......-M-A-M (MAPs-M) chain lying on the $z$-axis. Hence the MAP solutions possess zero net magnetic charge whereas the MAC solutions have a net magnetic charge of $n=1, 2$, \cite{kn:8}. The exact MAP asymptotic solutions of winding numbers $m$ and $n$ at small and large $r$ are \cite{kn:6}-\cite{kn:8}
\begin{eqnarray}
\psi_1 &=& \psi_2 = R_1 = R_2 = 0, \nonumber\\
\chi_1 &=& \xi_0 r\cos m\theta, ~~~~~\chi_2 = -\xi_0 r \sin m\theta,~~~\mbox{and}
\label{eq.13}
\end{eqnarray}
\begin{eqnarray}
\psi_1(r, \theta)&=& m, ~~~\psi_2(r, \theta)=\frac{n\sin m\theta}{\sin\theta},  \nonumber\\
R_1(r, \theta)&=&0, ~~~R_2(r, \theta)=\frac{n(\cos m\theta-1)}{\sin\theta}, \nonumber\\
\chi_1(r, \theta)&=& h+ \xi r ,~~~\chi_2(r, \theta)= 0, 
\label{eq.14}
\end{eqnarray}
respectively, where $h$, $\xi_0$, $\xi$ are constants and $m$ is an even integer. In this case, $\xi=\frac{\mu}{\sqrt{\lambda}}$ is the vacuum expectation value. In the BPS limit $h=0$, otherwise it is arbitrary. In Ref. \cite{kn:6}-\cite{kn:8}, numerical solutions were constructed at intermediate values of $r$ to join up the exact asymptotic soultions at large and small $r$. When $n=1$, $m=2$, the solution is a 1-MAP (M-A), when $n=1$, $m=4$, the solution is a 2-MAP (M-A-M-A), and so on.

The exact MAC asymptotic solutions at small and large $r$ are \cite{kn:6}-\cite{kn:8} are
\begin{eqnarray}
\psi_1 = \psi_2 = R_1 = R_2 = \chi_1 = \chi_2 = 0, ~~~\mbox{and}
\label{eq.15}
\end{eqnarray}
\begin{eqnarray}
\psi_1(r, \theta)&=& m, ~~~\psi_2(r, \theta)=\frac{n\sin m\theta}{\sin\theta},  \nonumber\\
R_1(r, \theta)&=&0, ~~~R_2(r, \theta)=\frac{n(\cos m\theta -\cos\theta)}{\sin\theta}, \nonumber\\
\chi_1(r, \theta)&=& h+\xi r ,~~~\chi_2(r, \theta)= 0, 
\label{eq.16}
\end{eqnarray}
respectively, where $m$ is an odd integer. In the BPS limit $h=n$ but it is however arbitrary outside this limit. Similarly, numerical solutions were obtained for intermediate values of $r$ in Ref. \cite{kn:6}-\cite{kn:8}. When $n=1$, $m=1$, the solution is the 't Hooft-Polyakov monopole, when $n=1$, $m=3$, the solution is a M-A-M chain, when $n=1$, $m=5$, the solution is a M-A-M-A-M chain,
and so on.

By using the relationship of the isospin unit vectors of winding numbers $m$ and $n$ with the isospin unit vectors of winding numbers $m=1$ and $n$, 
\begin{eqnarray}
\hat{u}^a_r &=& \cos(m-1)\theta ~\hat{n}_r^a + \sin(m-1)\theta ~\hat{n}_\theta^a, \nonumber\\
\hat{u}^a_\theta &=& -\sin(m-1)\theta ~\hat{n}_r^a + \cos(m-1)\theta ~\hat{n}_\theta^a,
\label{eq.17}
\end{eqnarray}
where
$\hat{n}_r^a = \sin \theta ~\cos n\phi ~\delta_{1}^a + \sin \theta ~\sin n\phi ~\delta_{2}^a + \cos \theta~\delta_{3}^a,~~
\hat{n}_\theta^a = \cos \theta ~\cos n\phi ~\delta_{1}^a + \cos \theta ~\sin n\phi ~\delta_{2}^a - \sin \theta ~\delta_{3}^a$, ~
we can reduced the asymptotic MAP and MAC solutions of Eq. (\ref{eq.13}) - (\ref{eq.16}) to solutions with winding number $m=1$ and a constant parameter $s$. Hence the MAP asymptotic solutions with winding numbers $n$, $m$ and integer parameter $s$ or $(n,m,s)$ are given by
\begin{eqnarray}
\psi_1 &=& \psi_2 = R_1 = R_2 = 0, \nonumber\\
\chi_1 &=& \xi_0 r\cos \theta, ~~~~~\chi_2 = -\xi_0 r \sin \theta,~~r \rightarrow 0 ~~\mbox{and}
\label{eq.18}
\end{eqnarray}
\begin{eqnarray}
\psi_1&=& 2s, ~~~\psi_2 = \frac{n(\sin m\theta + \sin(2s-m)\theta)}{\sin\theta}, \nonumber\\
R_1 &=&0, ~~~R_2 = \frac{n(\cos m\theta - \cos(2s-m)\theta)}{\sin\theta}, \nonumber\\
\chi_1&=& \left(h + \xi r\right)\cos(2s-m)\theta,~~~\chi_2=\left(h + \xi r\right)\sin(2s-m)\theta,~~r \rightarrow \infty.
\label{eq.19}
\end{eqnarray}
When $(n,m,s)=(1,1,0)$ in Eq. (\ref{eq.19}), the solution is just the trivial YMH vacuum. In fact the MAP asymptotic solution Eq. (\ref{eq.19}) is a pure gauge vacuum solution with zero net magnetic charge. Hence we expect to get a 1-MAP (M-A) solution when $(n,m,s)=(1,1,1)$, a 2-MAP (M-A-M-A) solution when $(n,m,s)=(1,1,2)$, and so on. In general Eq. (\ref{eq.19}) is a $s$-MAP solution when $(n,m,s)=(1,1,s)$.

Similarly the MAC asymptotic solutions with $(n,m,s)$ is given by
\begin{eqnarray}
\psi_1 = \psi_2 = R_1 = R_2=\chi_1=\chi_2 = 0, ~~r \rightarrow 0 ~~\mbox{and}
\label{eq.20}
\end{eqnarray}
\begin{eqnarray}
\psi_1&=& m+2s, ~~~\psi_2 = \frac{n(\sin m\theta + \sin 2s\theta \cos\theta)}{\sin\theta}, \nonumber\\
R_1&=&0, ~~~R_2=\frac{n(\cos m\theta - \cos 2s\theta\cos\theta)}{\sin\theta}, \nonumber\\
\chi_1&=& \left(h + \xi r\right)\cos2s\theta,~~~\chi_2=\left(h + \xi r\right)\sin2s\theta,~~r \rightarrow \infty.
\label{eq.21}
\end{eqnarray}
When $(n,m,s)=(1,1,0)$  in Eq. (\ref{eq.21}), we get the 't Hooft-Polyakov monopole solution which is just a finite energy one monopole solution \cite{kn:1}, \cite{kn:2}. In the BPS limit, we get the exact one monopole solution of Ref. \cite{kn:2}. When $(n,m,s)=(1,1,1)$ the MAC solution is a M-A-M chain, whereas when $(n,m,s)=(1,1,2)$ we get the M-A-M-A-M chain of monopoles, and so on.

In the work of Kleihaus et al. \cite{kn:7}-\cite{kn:8}, the accuracies of the numerical computation decreases as the winding number $m$ increases.
The purpose of reducing the MAP and MAC solutions to the $m=1$ winding number form of the solution is that we can retain the accuracies of the numerical computation at winding number $m=1$ as the parameter $s$ varies. This is particularly useful when we are considering the higher pairs and chains solutions. For better accuracies, we can compute the numerical $m=1$ form instead with lower error of the order of $\times 10^{-4}$.

%when we compute the numerical 1-MAP solution for intermediate values of $r$ connecting the two asymptotic solutions Eq. (\ref{eq.18}) and (\ref{eq.19}) at small and large $r$ respectively, we notice that when we increased the accuracy of the numerical solutions, the position of the pole of the magnetic dipole on the positive $z$-axis approaches a critical point on its left from the right when $(n,m,s)=(1,1,1)$. In the case when $(n,m,s)=(1,2,0)$, the position of the magnetic pole approaches the critical point on its right from the left as the accuracy increases. If the exact connecting solution is unique, then the position of the magnetic pole lies on the critcal point in between these two sets of numerical solutions. Hence by running both $(n,m,s)=(1,1,1)$ and $(n,m,s)=(1,2,0)$ solutions will give a more accurate location of the magnetic monopoles. 

%%%%%%%%%%%%%%%%%%%%%%%%%% The exact BPS One Monopole %%%%%%%%%%%%%%%%%%%%%%%%%%%%%%

\section{The exact BPS One Monopole}

\noindent To solve for exact solutions, the ansatz (\ref{eq.10}) is substituted into the Bogomol'nyi equation (\ref{eq.4}). We find that the BPS exact one monopole solution with higher $\theta$-winding number is 
\begin{eqnarray}
\psi_1(r, \theta)&=& m\pm \frac{\zeta r}{\sinh \zeta r}, ~~~\psi_2(r, \theta)=\frac{n\sin m\theta}{\sin\theta} \pm \frac{\zeta r}{\sinh \zeta r}, \nonumber\\
R_1(r, \theta)&=&0, ~~~R_2(r, \theta)=\frac{n\cos m\theta-\cos\theta}{\sin\theta}, \nonumber\\
\chi_1(r, \theta)&=& 1-\frac{\zeta r}{\tanh \zeta r},~~~\chi_2(r, \theta)=0.
\label{eq.22}
\end{eqnarray}
where $\zeta$ is a constant. The solution (\ref{eq.22}) is non singular only when $r>0$, $m$ is odd and $n=1$. Although the solution is singular at $r=0$ for $m>1$, the magnetic field and the energy density are finite and bounded.
Upon calculating for the non-Abelian magnetic field of solution (\ref{eq.22}), we get
\begin{eqnarray}
B^a_i =  \frac{1}{r^2}\left\{\left(1-\frac{\zeta^2 r^2}{\sinh^2 \zeta r}\right)\hat{u}^{a}_r\hat{r}_i - \frac{\zeta r}{\sinh \zeta r}\left(1-\frac{\zeta r}{\tanh \zeta r}\right)(\hat{u}^{a}_\theta\hat{\theta}_i + \hat{u}^{a}_\phi\hat{\phi}_i)\right\},
\label{eq.23}
\end{eqnarray}
which is just the BPS one monopole magnetic field of higher winding number $m$. The  magnitude of the magnetic field, $B^a_i$, is independent of the value of $m$. However, they possess different isopin direction in internal space for different values of $m$.
The energy or mass of the system is finite,
\begin{eqnarray}
{\cal E} = \int~dx^3 (\frac{1}{4}F^a_{ij}F^a_{ij} + \frac{1}{2}F^a_{0i}F^a_{0i} + \frac{1}{2}D_i\Phi^a D_i\Phi^a + \frac{1}{2}D_0\Phi^a D_0\Phi^a)
 = 4\pi\zeta.
\label{eq.24}
\end{eqnarray}

\noindent The Abelian gauge potential $A_\mu$ is non vanishing and is given by 
\begin{equation}
A_\mu=\left\{\frac{\cos\theta-\cos m\theta}{r\sin\theta}\right\}\hat{\phi}_\mu.
\label{eq.25}
\end{equation}
Hence the gauge part of the Abelian magnetic field is
\begin{equation}
B_i^G=\left(1-\frac{m\sin m\theta}{\sin\theta}\right)\frac{\hat{r}_i}{r^2},
\label{eq.26}
\end{equation}
which when $m=3$ is the magnetic field of a zero length A-M-M-A chain at the origin along the $z$-axis. Hence the net magnetic charge is zero. The Higgs part of the Abelian magnetic field is
\begin{equation}
B_i^H=\frac{m\sin m\theta}{\sin\theta}\frac{\hat{r}_i}{r^2},
\label{eq.27}
\end{equation}
and it corresponds to a M-A-M chain of zero length at the origin of net magnetic charge one. 

By using Eq. (\ref{eq.17}), we can reduced the winding number $m$ of solution (\ref{eq.22}) to one. Hence solution (\ref{eq.22}) with $m=1$ and $n=1$ is given by
\begin{eqnarray}
\psi_1&=& 1+2s\pm \frac{\zeta r}{\sinh\zeta r}, ~~~\psi_2 = \frac{\sin\theta + \sin2s\theta \cos\theta}{\sin\theta} \pm \frac{\zeta r \cos2s\theta}{\sinh\zeta r}, \nonumber\\
R_1&=&0, ~~~R_2=\frac{\cos\theta - \cos2s\theta\cos\theta}{\sin\theta}\pm \frac{\zeta r \sin2s\theta}{\sinh\zeta r}, \nonumber\\
\chi_1&=& \left(1-\frac{\zeta r}{\tanh\zeta r}\right)\cos2s\theta,~~~\chi_2=\left(1-\frac{\zeta r}{\tanh\zeta r}\right)\sin2s\theta.
\label{eq.28}
\end{eqnarray}
Solution (\ref{eq.28}) is equivalent to solution (\ref{eq.22}) of winding number $m=2s+1$, where $s$ a natual number.
The exact BPS one monopole solution of Ref. \cite{kn:2} corresponds to solution (\ref{eq.28}) when $s=0$. When $s=1$, the solution is
\begin{eqnarray}
\psi_1 &=& 3 \pm \frac{\zeta r}{\sinh\zeta r}, ~~~\psi_2 = 2 + \cos2\theta\left(1 \pm \frac{\zeta r}{\sinh\zeta r}\right), \nonumber\\
R_1 &=&0, ~~~R_2= \sin2\theta \left(1 \pm \frac{\zeta r}{\sinh\zeta r}\right), \nonumber\\
\chi_1&=& \left(1-\frac{\zeta r}{\tanh\zeta r}\right)\cos2\theta,~~~\chi_2=\left(1-\frac{\zeta r}{\tanh\zeta r}\right)\sin2\theta.
\label{eq.29}
\end{eqnarray}
This general exact one monopole BPS solution (\ref{eq.29}) possess finite energy. It is non singular at all space except at the origin when $r$ tends to zero. However at small $r$, the gauge field is a pure gauge with $B^a_i=0$.

%%%%%%%%%%%%%%%%%%%%%% The Numerical Solutions %%%%%%%%%%%%%%%
\section{The Numerical Solutions}
With ansatz (\ref{eq.10}) the field equations (\ref{eq.3}) reduce to six partial differential equations in $r$ and $\theta$. The ansatz possesses a residual U(1) gauge symmetry. In order to fix the gauge, we impose the gauge condition $r\frac{\partial}{\partial r}R_1-\frac{\partial}{\partial \theta}\psi_1$. 

In Refs. \cite{kn:6}-\cite{kn:7}, Kleihaus et al. consider MAP and MAC solutions which is parameterized by the $\theta$-winding number $m$ ($>1$) and  $\phi$-winding number $n$ ($=1$). In Section 3 \cite{kn:9}, we show that there always exist an equivalent form of the solutions with normal winding number $m = 1$ and an integer $s$ for all the solutions with $\theta$-winding number $m>1$. 

The MAP solution possesses exact asymptotic solutions at both small and large $r$. Upon reducing these solutions to the $m = 1$ form with integer $s$, both the MAP and MAC solutions correspond to the trivial vacuum ($s = 0$) at small $r$. However, at large $r$, the MAP solutions tend to a different sector of the vacuum with parameter $s = 1, 2, 3$... Hence the MAP solutions correspond to a one monopole-antimonopole pair when $(n,m,s)=(1,1,1)$ and also when $(n,m,s)=(1,2,0)$. Here we compute numerically the MAP solution with $(n,m,s)=(1,1,1)$ and $(n,m,s)=(1,2,0)$ at different numerical accuracies \cite{kn:10}. 

The boundary conditions at the origin of the $(n,m,s)=(1,1,1)$ MAP solution are
\begin{eqnarray}
&&\psi_1(0, \theta)\rightarrow 0, ~~ \psi_2(0, \theta)\rightarrow 0, ~~ R_1(0, \theta)\rightarrow 0, ~~ R_2(0, \theta)\rightarrow 0, ~~ \nonumber\\
&&\Phi_1(0, \theta)\rightarrow -\xi_0 \cos\theta, ~~ \Phi_2(0, \theta)\rightarrow \xi_0 \sin\theta,
\label{eq.30}\\
&&\sin\theta~\Phi_1(0, \theta)+\cos\theta~\Phi_2(0, \theta)=0, ~~ \nonumber\\
&&\frac{\partial}{\partial r}\left(\cos\theta~\Phi_1(r, \theta)-\sin\theta~\Phi_2(r, \theta)\right)|_{r=0}=0,
\label{eq.31}
\end{eqnarray}
where $\xi_0$ is an arbitrary constant and its boundary conditions at $r$ infinity are
\begin{eqnarray}
\psi_1(\infty, \theta)&\rightarrow& 2, ~~ \psi_2(\infty, \theta)\rightarrow 2, ~~ R_1(\infty, \theta)\rightarrow 0, ~~ R_2(\infty, \theta)\rightarrow 0, ~~ \nonumber\\
\Phi_1(\infty, \theta)&\rightarrow& \xi \cos\theta, ~~ \Phi_2(\infty, \theta)\rightarrow \xi \sin\theta
\label{eq.32}
\end{eqnarray}
Regularity on the $z$-axis (at $\theta=0$ and $\theta=\pi$) requires 
\begin{eqnarray}
R_1 = R_2 = \Phi_2 = 0, ~~ \frac{\partial}{\partial \theta}\psi_1 = \frac{\partial}{\partial \theta}\psi_2 = \frac{\partial}{\partial \theta}\Phi_1 = 0. 
\label{eq.33}
\end{eqnarray}
                                              
The equations of motion (\ref{eq.3}) are then solved numerically by using ansatz (\ref{eq.10}) with the boundary conditions (\ref{eq.30})-(\ref{eq.33}). The constant $\xi_0$ will be determined by the results of the numerical calculations.

To calculate for the Abelian magnetic field $B_i$, we rewrite the Higgs field in Eq. (\ref{eq.10}) from the spherical to the Cartesian coordinate system, \cite{kn:6}, \cite{kn:7}.
\begin{eqnarray}
\Phi^a &=& \Phi_{1}~\hat{r}^{a} + \Phi_{2}~\hat{\theta}^{a} + \Phi_3~\hat{\phi}^{a}\nonumber\\
&=& \tilde{\Phi}_1 ~\delta^{a1} + \tilde{\Phi}_2 ~\delta^{a2} + \tilde{\Phi}_3 ~\delta^{a3}
\label{eq.34}
\end{eqnarray}
\begin{eqnarray}
\mbox{where}~~~\tilde{\Phi}_1 &=& \sin m\theta \cos n\phi ~\Phi_1 + \cos m\theta \cos n\phi ~\Phi_2 - \sin n\phi ~\Phi_3\nonumber\\
&=& |\Phi|\cos\alpha \sin\beta\nonumber\\
\tilde{\Phi}_2 &=& \sin m\theta \sin n\phi ~\Phi_1 + \cos m\theta \sin n\phi ~\Phi_2 + \cos n\phi ~\Phi_3\nonumber\\
&=& |\Phi|\cos\alpha \cos\beta\nonumber\\
\tilde{\Phi}_3 &=& \cos m\theta ~\Phi_1 - \sin m\theta ~\Phi_2 = |\Phi|\sin\alpha,
\label{eq.35}
\end{eqnarray}
The Higgs unit vector is then simplified to 
\begin{equation}
\hat{\Phi}^a = \cos\alpha \sin\beta ~\delta^{a1} + \cos\alpha \cos\beta ~\delta^{a2} + \sin\alpha ~\delta^{a3}
\label{eq.36}
\end{equation}
where
\begin{eqnarray}
\sin\alpha &=& \frac{\Phi_1\cos m\theta - \Phi_2 \sin m\theta}{\sqrt{\Phi_1^2+\Phi_2^2}}, ~~~\beta = \frac{\pi}{2} - n\phi,
\label{eq.37}
\end{eqnarray}
and the Abelian magnetic field is found to be
\begin{eqnarray}
B_i &=& \frac{1}{r^2 \sin\theta}\left\{\frac{\partial\sin\alpha}{\partial\theta}\frac{\partial\beta}{\partial\phi} - \frac{\partial\sin\alpha}{\partial\phi}\frac{\partial\beta}{\partial\theta}\right\}\hat{r}_i  \nonumber\\
&+& \frac{1}{r\sin\theta}\left\{\frac{\partial\sin\alpha}{\partial\phi}\frac{\partial\beta}{\partial r} - \frac{\partial\sin\alpha}{\partial r}\frac{\partial\beta}{\partial\phi}\right\}\hat{\theta}_i.
\label{eq.38}
\end{eqnarray}
 		                                         
The topological magnetic charge is defined as in Eq. (\ref{eq.9}). Here we define the magnetic charge enclosed by the upper hemisphere of infinite radius as $M_+$ whereas the magnetic charge enclosed by the lower hemisphere of infinite radius is denoted by $M_-$. The value of $M_+$ is calculated to be 
\begin{eqnarray}
M_+ = -\left.\frac{1}{2}\sin\alpha\right|^{\pi/2}_{0, r\rightarrow\infty} = +1,
\label{eq.39}
\end{eqnarray}                                                                                 
when $n=1$. The upper hemisphere then possesses a positive charged magnetic monopole. Similarly when $n=1$,
\begin{eqnarray}
M_- = -\left.\frac{1}{2}\sin\alpha\right|^{\pi}_{\pi/2, r\rightarrow\infty} = -1.
\label{eq.40}
\end{eqnarray}  
which correspond to a negative magnetic charged antimonopole. 
These calculations show indeed that the configuration possesses a monopole-antimonopole pair, with the monopole situated on the positive $z$-axis and the antimonopole at equidistance on the negative $z$-axis. At $r$ infinity for surface enclosing both charges, their contributions compensate and yields zero net magnetic charge. 

The asymptotic conditions (\ref{eq.32}) at large $r$ correspond to a pure gauge vacuum at spatial infinity. We connect this asymptotic condition with the trivial vacuum at the origin (\ref{eq.30}) numerically by using mathematical softwares Maple and Matlab.

\begin{figure}[tbh]
	\centering
		\includegraphics[scale=0.24]{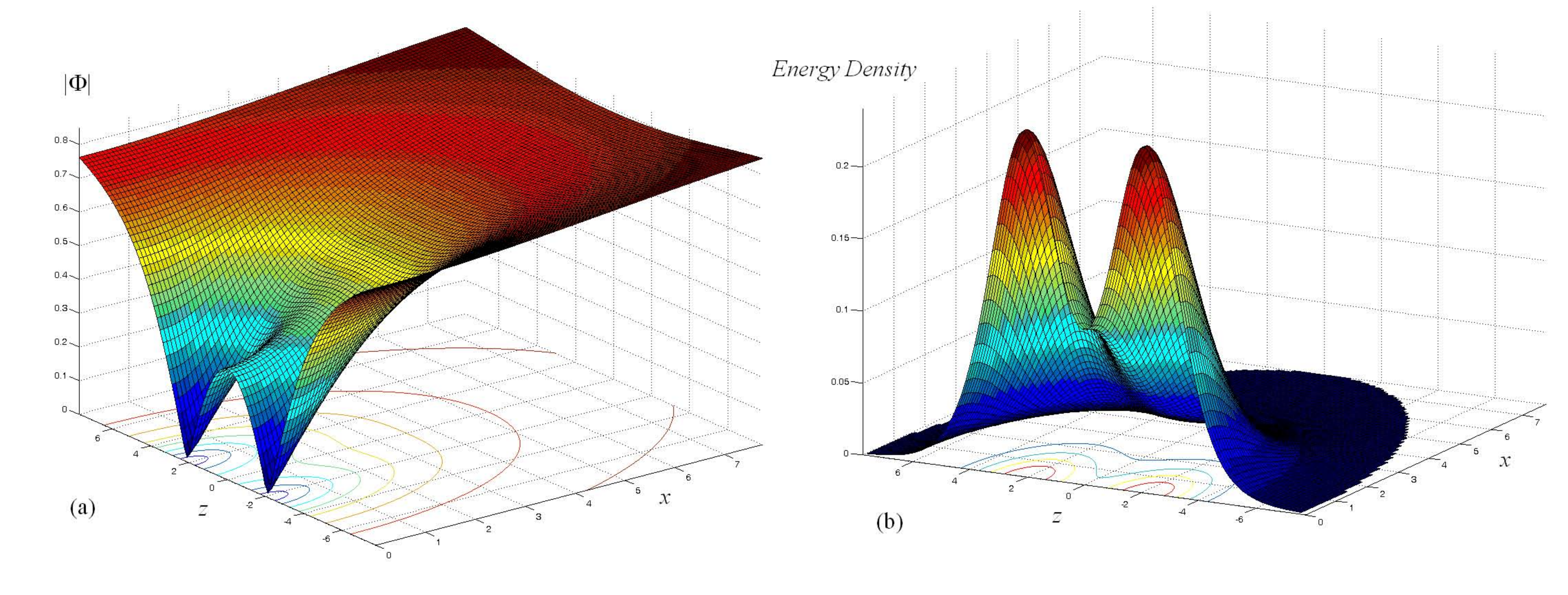}
	\caption{Three dimensional plot of the (a) magnitude of the Higgs field $|\Phi$ and (b) energy density of the $(n,m,s)=(1,1,1)$ MAP solution at $(M, N) = (70, 60)$.}
	\label{fig.1}
\end{figure}
%\begin{figure}[tbh]
%\vspace{3.0in}
%\hskip-0.1in\special{bmp:Higgs_ED.bmp x=6in y=3.0in}
%\caption{Three dimensional plot of the (a) magnitude of the Higgs field $|\Phi$ and (b) energy density of the $(n,m,s)=(1,1,1)$ MAP solution at $(M, N) %= (70, 60)$.}
%\label{fig.1}
%\end{figure} 

The equations of motion are transformed into a system of nonlinear  equations by using finite difference approximation and discretized on a non-equidistant grid, covering the integration region $0 \leq \bar{x} \leq 1$ and $0 \leq \theta \leq \pi$. Here $\bar{x}=\frac{r}{1+r}$ is the finite interval compactified coordinate. The partial derivative with respect to the radial coordinate is then replaced accordingly by
$\partial_r \rightarrow (1-\bar{x})^2 \partial_{\bar{x}}$ and ~~$\frac{\partial^2}{\partial r^2} \rightarrow (1-\bar{x})^4\frac{\partial^2}{\partial \bar{x}^2} - 2(1-\bar{x})^3\frac{\partial}{\partial \bar{x}}$.
The $\bar{x}$ and $\theta$ grid are subdivided into $M$ and $N$ divisions. The best accuracy our computer is able to support is $M = 70$ and $N = 60$. 
The numerical method used is the Gauss-Newton method and it is a good iterative method to obtain numerically accurate solutions. After providing good initial guess to the system of nonlinear equations, the solver will iterate and converge to the true numerical answers. 
Our result confirms that the boundary conditions (\ref{eq.30})-(\ref{eq.33}) corresponds to monopole-antimonopole pair solution. We solved numerically for the $(n,m,s)=(1,1,1)$ and $(n,m,s)=(1,2,0)$
 MAP solutions at different accuracies when $(M, N) = (50, 40), ~(60, 50)$ and $(70, 60)$. The monopoles' location $\pm z_0$ along the $z$-axis, the total energy $E$ of the system, and the Higgs field magnitude at the origin, $\xi_0$ are calculated at the three different accuracies. The results are as given in Table \ref{table:1} and \ref{table:2}.~~\\

\begin{table}[tbh]
\begin{tabular}{|c|c|c|c|} \hline
~~~\bf{$(M, N)$}~~~		&~~~~$(50, 40)$~~	&~~~~$(60, 50)$~~	&~~~~$(70, 60)$ 	\\ \hline
Monopole's location, $\pm z_0$  & ~$\pm 2.105$     	& ~$\pm 2.101$         	& ~$\pm 2.098$ 		\\ \hline
Total energy, $E$            	& ~1.6952     		& ~1.6938      		& ~1.6934         	\\ \hline
Higgs magnitude, $\xi_0$        & ~0.3248    		& ~0.3258        	& ~0.3264      		\\ \hline
\end{tabular}
\caption{Some of the $(n,m,s)=(1,1,1)$ MAP solution properties calculated at different accuracies.}
\label{table:1}
\end{table}
\begin{table}[tbh]
\begin{tabular}{|c|c|c|c|} \hline
~~~\bf{$(M, N)$}~~~		&~~~~$(50, 40)$~~	&~~~~$(60, 50)$~~	&~~~~$(70, 60)$ 	\\ \hline
Monopole's location, $\pm z_0$  & ~$\pm 2.081$    	& ~$\pm 2.085$         	& ~$\pm 2.087$ 		\\ \hline
Total energy, $E$            	& ~1.6969     		& ~1.6905      		& ~1.6942         	\\ \hline
Higgs magnitude, $\xi_0$        & ~0.2897     		& ~0.3018        	& ~0.3089      		\\ \hline
\end{tabular}
\caption{Some of the $(n,m,s)=(1,2,0)$ MAP solution properties calculated at different accuracies.}
\label{table:2}
\end{table}

In Fig.(\ref{fig.1}) we plot the energy density of the $(n,m,s)=(1,1,1)$ MAP solution as a functions of $\rho$ and $z$ where $\rho=\sqrt{x^2+y^2}$ at the accuracy of $(M,N) = (70,60)$. 
The magnetic poles are located at the peaks of the energy density plot. %A plot the Higgs field strength along the $z$-axis is shown in Fig. (\ref{fig.2}) for the $(n,m,s)=(1,1,1)$ and $(n,m,s)=(1,2,0)$ MAP solutions.

\begin{figure}[tbh]
	\centering
		\includegraphics[scale=0.6]{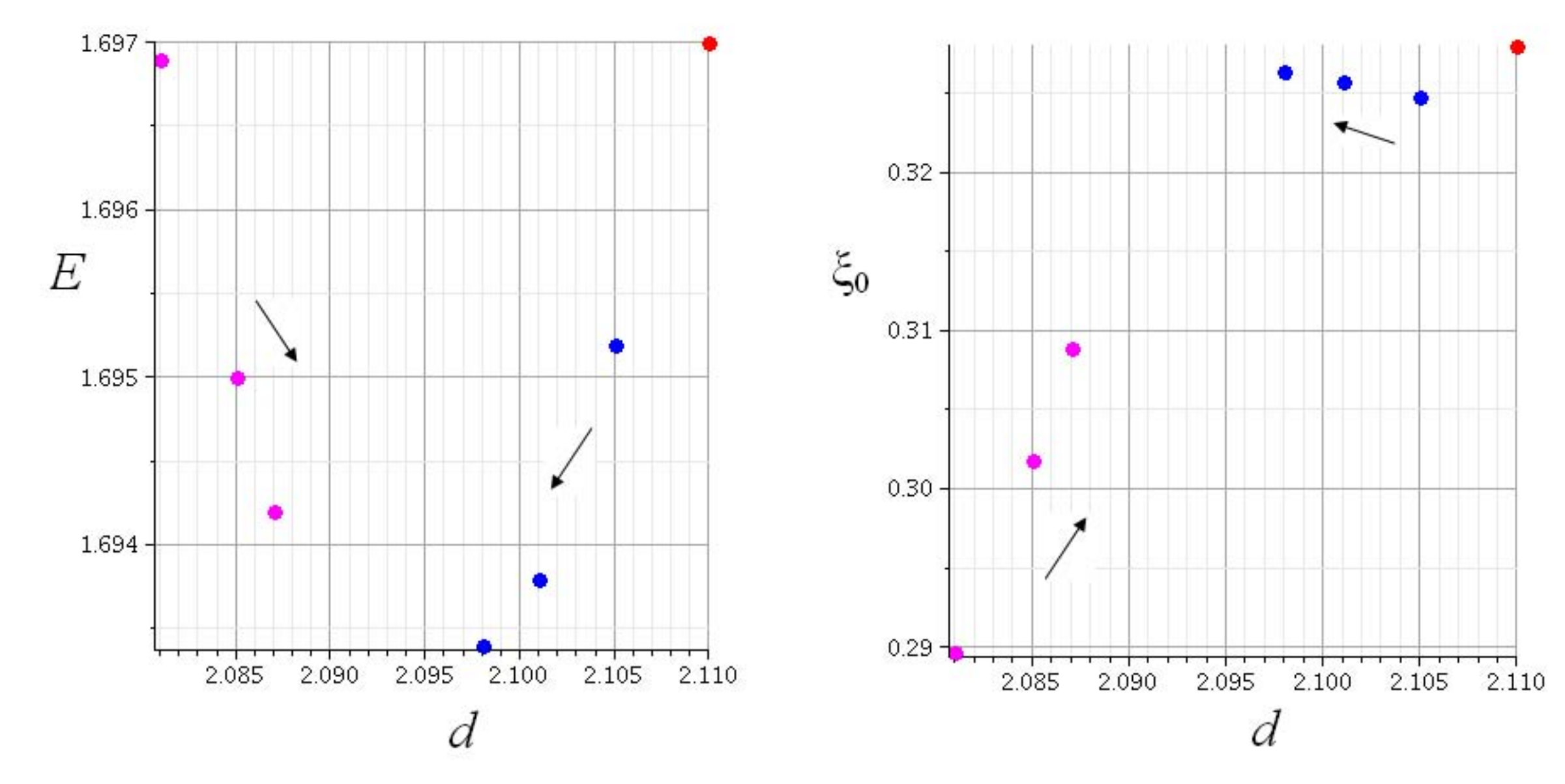}
	\caption{Point plots the total energy $E$ and Higgs field magnitude, $\xi_0$, versus the poles separation $d$ along the $z$-axis for the $(n,m,s)=(1,1,1)$ (blue), $(n,m,s)=(1,2,0)$ (purple) and Kleihaus-Kunz (red) MAP solutions at various accuracies. The arrows show the direction of increasing accuracy.}
	\label{fig.2}
\end{figure}
%\begin{figure}[tbh]
%\vspace{3.0in}
%\hskip-0.2in\special{bmp:E_xi0_d.bmp x=6.0in y=3.0in}
%\caption{Point plots the total energy $E$ and Higgs field magnitude, $\xi_0$, versus the poles separation $d$ along the $z$-axis for the $(n,m,s)=(1,1,1)$ (blue), $(n,m,s)=(1,2,0)$ (purple) and Kleihaus-Kunz (red) MAP solutions at various accuracies. The arrows show the direction of increasing accuracy.}
%\label{fig.2}
%\end{figure} 

From the numerical solutions we have notice that with increasing accuracy, the $(n,m,s)=(1,2,0)$ MAP plots of Fig. (\ref{fig.2}) shift to the right whereas the $(n,m,s)=(1,1,1)$ MAP plots shift to the left with both sets approaching a critical value (not drawn) for both the total energy $E$ and Higgs field magnitude $\xi_0$ versus poles separation $d$ point plots. Since the $(n,m,s)=(1,2,0)$ and the $(n,m,s)=(1,1,1)$ MAP solutions are practically the same solution, there must exist only one critical value for $E$, $\xi_0$, and $d$. Hence our work have shown that in order to get the best accuracy for the properties of the MAP solution, we should take the average value of the two sets of solutions \cite{kn:10} with different $m$..

%%%%%%%%%%%%%%%%%%% The Comments %%%%%%%%%%%%%%%%%%%%%%%%%%%
\section{Comments}
The MAP and MAC are finite energy monopoles chain solutions of net magnetic charge zero and $n\leq 2$ respectively. As the exact form of these solutions have not yet been found, highly accurate numerical solutions will be useful in studying their properties. 

We have shown that all the MAP and MAC solutions obtained from the equations of motions (\ref{eq.3}) with higher $\theta$-winding number $m>1$, can always be reduced to the $m=1$ form. Analytically both forms represent the same physical solutions. However numerically they give different accuracies when performed with the same numerical procedures and mathematical softwares. For the $(n,m,s)=(1,2,0)$ and the $(n,m,s)=(1,1,1)$ MAP solutions, we have shown that the Higgs field strength curves approach a critical curve, Fig. (\ref{fig.2}). This is useful in getting a good estimated value for the properties of the MAP solution since the analytic solution has not yet been found. We have also noted that by computing these numerical solutions in the $m=1$ $\theta$-winding number forms do give more accurate results. 

For a comparison with previous work on the MAP solution \cite{kn:6} - \cite{kn:8}, the values of the monopole's location, $\pm z_0$, total energy, $E$, and Higgs magnitude, $\xi_0$ obtained by these aurthors are $\pm 2.11$, 1.697, and 0.328. However these values are obtained by using different mathematical softwares.

We have also shown that the exact BPS one monopole solution has a pure gauge vacuum $s$ solution (not necessarily the trival vaccum) at small $r$ and a Wu-Yang type monopole $s$ solution at large $r$ and hence is just a one monopole solution.

We have also noticed that by writing the MAP and MAC solutions in $m=1$ $\theta$-winding number form, we are able to find more general exact asymptotic solutions at large distances. These generalized asymptotic solutions which are Jacobi Elliptic in nature leads to new MAP solutions which are axially symmetic \cite{kn:13}. We also report on a new axially symmetric one-monopole in Ref. \cite{kn:14}. In Ref. \cite{kn:15}, more new axially symmetric one-monopole configurations will be discussed.

%%%%%%%%%%%%%%%%%%%%% Acknowlegements %%%%%%%%%%%%%%%%%%%%%%%%%%%%%

\section*{Acknowlegements}
The authors would like to thank the Ministry of Science, Technology and Innovation (MOSTI) of Malaysia for the award of ScienceFund research grant (Project Number: 06-01-05-SF0266) and Universiti Sains Malaysia for financial support.

\newpage

\end{document}